\documentclass[10pt]{iopart}

\usepackage{bm}
\usepackage{graphicx}

\begin{document}

\title{Defect kinetics and dynamics of pattern coarsening in a two-dimensional smectic-A system}

\author{Nasser Mohieddin Abukhdeir and Alejandro D Rey}
\address{Department of Chemical Engineering, McGill University, Montreal, Quebec, Canada H3A 2B2}
\eads{\mailto{nasser.abukhdeir@mcgill.ca}, \mailto{alejandro.rey@mcgill.ca}}

\begin{abstract}
Two-dimensional simulations of the coarsening process of the isotropic/smectic-A phase transition are presented using a high-order Landau-de Gennes type free energy model.  Defect annihilation laws for smectic disclinations, elementary dislocations, and total dislocation content are determined.  The computed evolution of the orientational correlation length and disclination density is found to be in agreement with previous experimental observations showing that disclination interactions dominate the coarsening process.  The mechanism of smectic disclination movement, limited by the absorption and emission of elementary dislocations, is found to be facilitated by curvature walls connecting interacting disclinations.  At intermediate times in the coarsening process, split-core dislocation formation and interactions displaying an effective disclination quadrupole configuration are observed.  This work provides the framework for further understanding of the formation and dynamics of the diverse set of curvature defects observed in smectic liquid crystals and other layered material systems.
\end{abstract}

\pacs{42.70.Df, 61.30.Jf, 64.70.M-, 07.05.Tp}
\submitto{\NJP}

\maketitle

\tableofcontents

\section{Introduction}
The study of liquid crystalline mesophases has resulted in profound contributions, both through innovative technological advancements and through the increase of fundamental understanding.  Examples of technological applications range from display technology to the cutting-edge field of responsive materials \cite{Mather2007}.  Unlike orientationally ordered nematic liquid crystals, higher order mesophases have garnered less attention.  Lamellar and columnar liquid crystals, with their degree of translational ordering, open up an exciting avenue of progress.  For instance, the study of columnar mesophases could shed light on the early formation of life on Earth \cite{Nakata2007}.  Focusing on the simplest of these higher ordered mesophases, the smectic-A mesophase is a physical system which serves as a template for striped systems in general.  Striped systems are present in a diverse number of areas ranging from biological membranes to flexible polymer crystallization \cite{Li2003}.

Much excitement regarding the smectic-A mesophase is due to its sharing of symmetries with certain types of self-assembled block copolymer films.  This has recently been shown by Harrison et al through experimental observation \cite{Harrison2000,Harrison2002}.  Given the possible applications of these block copolymer systems in photolithography \cite{Ruiz2007,Stoykovich2007}, understanding of their kinetics and dynamics is of great importance.  Experimental work that has been focused in this direction is limited due to the time scales (nanoseconds) and length scales (nanometers) involved.  These limitations provide a relevant application for modelling and simulation approaches, which have been demonstrated as a powerful tool for the prediction of phase ordering behaviour in similar systems \cite{Duchs2002}.  Applications of these films to enhance current photolithographic techniques in the semiconductor industry are key to advancement in this field \cite{Black2007}.

In addition to industrial motivations, study of the isotropic/smectic-A mesophase transition is interesting for contributions to the expansion of fundamental phase transition knowledge.  Research cross-fertilization has played a key role in theoretical advances through the studying of small-scale rapid phase ordering transitions.  For example, within the frame of defect kinetics and dynamics, experiments and simulations of the isotropic/nematic mesophase transition have been used as a template for the study of the formation of the early universe \cite{Kibble2007}.  Additionally, a general topological theory of defects has been cultivated through the study of liquid crystalline mesophases as a basis \cite{Kleman2008}.  While defects of these mesomorphic phases are well understood from a macroscopic view, the microscopic structure and their interaction dynamics are not \cite{Kleman2008,Michel2006}.  In general, experimental methods are currently limited to static macroscopic studies of mesophase defects.  While in recent years experimental methods have begun to bridge this gap \cite{Harrison2000,Harrison2002,Ruiz2007,Michel2006}, theoretical approaches are vital to further study in this field.

This work, using modelling and simulation in two-dimensions, studies the first-order isotropic/smectic-A mesophase transition.  Model parameters were determined, in part, using experimental data from 12CB (dodecyl-cyanobiphenyl) \cite{Abukhdeir2007}.  Simulation of a temperature quench below the lower stability limit of the isotropic phase was performed.  The resulting coarsening kinetics, defect structure, and dynamics are the focus of this work.  The simulation results presented for the two-dimensional smectic-A system show strong agreement with experimental observations of a system with the same symmetries, self-assembled cylindrical block copolymer films \cite{Harrison2000,Harrison2002}. In addition to experimental agreement, simulation results show a mesoscopically complete view of the defect interaction process.  As explained in full detail below, we find that disclination interactions are facilitated by curvature walls connecting them.  These curvature walls are also found to be key in the formation of split-core dislocations where pairs of interacting disclinations of opposite sign form dislocations of high Burger's vector.  Quadrupolar configurations of disclinations are found to dominate defect interactions in the intermediate coarsening regime.  The presence of these disclination configurations result in interesting rheological properties of many different materials exhibiting smectic ordering \cite{Wood1986}.

This article is organized as follows: a brief background on relevant types of liquid crystalline order and defects (Sections \ref{secmesophases}-\ref{secdefects}) is given, the model and simulation approach are explained (\Sref{secmodel}), and simulation results are discussed (\Sref{secresanddisc}).  Two different viewpoints are used to understand the simulation results: i) overall defect/texture kinetics and ii) specific defect dynamics.  Overall defect/texture kinetics addresses multi-body interactions and the overall evolution of the texture of the liquid crystalline domain.  This type of analysis is most comparable to experimental observations of similar systems.  The final sections of this article address specific defect interactions at time and length scales that are not fully accessible experimentally.

\section{Background and theory} \label{secbackground}

\subsection{Liquid crystalline order} \label{secmesophases}

Liquid crystalline phases or mesophases are materials which exhibit partial orientational and/or translational order.  They are composed of anisotropic molecules  which can be disc-like or rod-like in shape.  Thermotropic liquid crystals are typically pure-component compounds that exhibit mesophase ordering most greatly in response to temperature changes.  Lyotropic liquid crystals are mixtures of mesogens, possibly with a solvent, that most greatly exhibit mesophase behaviour in response to concentration changes.  Effects of pressure and external fields also influence mesophase behaviour.  This work focuses the study of a rod-like thermotropic liquid crystals which exhibit a first-order liquid crystalline phase transition.

\begin{figure}[htp]
\begin{center}
\includegraphics[width=6in]{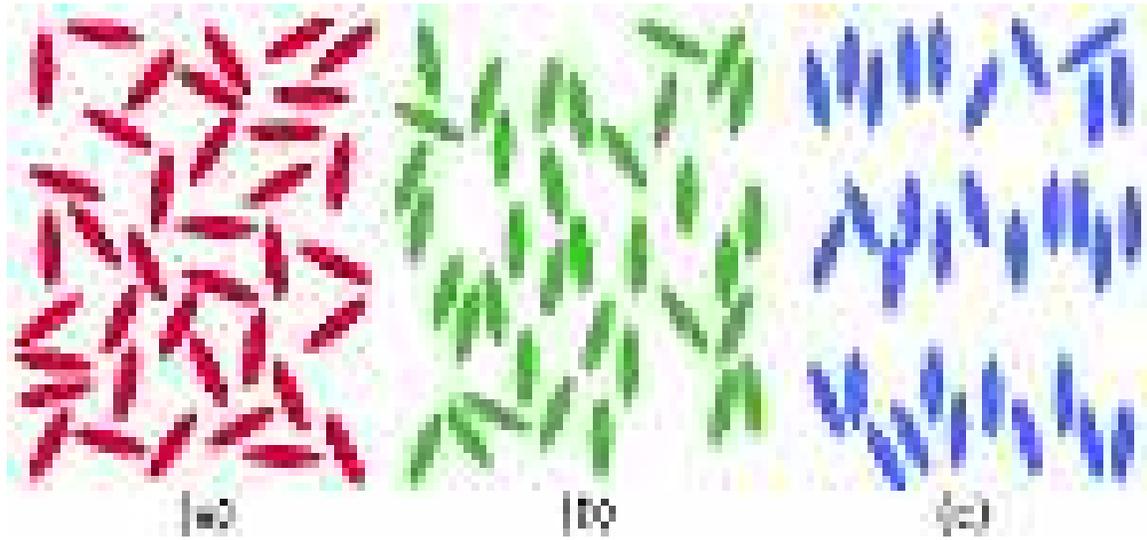} 
\end{center}
\caption{schematics of a) the isotropic, b) nematic, and c) smectic-A mesophases. \label{figlcorder}}
\end{figure}

An unordered liquid, where there is neither orientational nor translational order (apart from an average intermolecular separation distance) of the molecules, is referred to as isotropic.  Of the many different mesophases observed, the two of interest in this work are the nematic and smectic-A.  The nematic mesophase exhibits partial orientational order, where molecules conform to a local preferred orientational axis.  Smectic liquid crystals share the orientational order of nematics, but additionally have partial translational order.  The simplest of the smectics is the smectic-A mesophase, which exhibits one-dimensional lamellar order in the direction of the preferred molecular orientational axis.  It can be thought of as layers of two-dimensional fluids stacked upon each other.  A schematic representation of these three phases are shown in \Fref{figlcorder}.

Theoretical characterization of mesophase order is accomplished using order parameters that adequately capture the physics involved.  These order parameters typically have an amplitude and phase associated with them.  In order to characterize the partial orientational order of the nematic phase, a second order symmetric traceless tensor can be used:
\begin{equation} \label{eqnem_order_param}
\bi{Q} = S (\bi{nn} - \frac{1}{3} \bi{I})
\end{equation}
where $\mathbf{n}$ is a unit vector along the average orientational axis and $S$ is a scalar which represents the extent to which the molecules conform to the average orientational axis.  For simplicity, the expression for the tensor order parameter \eref{eqnem_order_param} is simplified using the assumption of uniaxial orientational order; refer to \cite{Wincure2007} for full treatment of the biaxial case.

The smectic-A mesophase has one-dimensional translational order in addition to the orientational order found in nematics.  Characterizing this mesophase can be accomplished through the use of primary (orientational) and secondary (translational) order parameters together \cite{Toledano1987}.  This is accomplished using the tensor order parameter \eref{eqnem_order_param} and the complex order parameter:
\begin{equation} \label{eqsmec_order_param}
\Psi = \psi e^{i \phi}
\end{equation}
where $\phi$ is the phase and $\psi$ is the scalar amplitude of the density modulation.

The focus of this work is on the isotropic/smectic-A mesophase transition.  Experimental observations of this transition show that it is first-order, where material properties and order parameters change discontinuously across the phase transition.  Examples of two of the more well characterized materials that exhibit this transition are 10CB and 12CB (cyanobiphenyl) \cite{Coles1979a}.

\begin{figure}[htp]
\begin{center}
\includegraphics[width=6in]{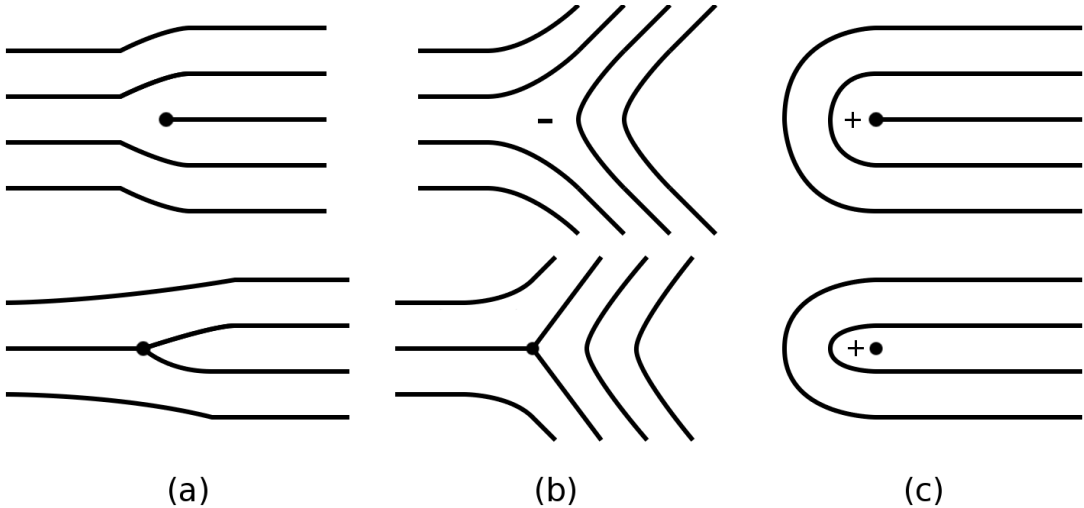} 
\end{center}

\caption{Schematics of different layer structures in the vicinity of a) an elementary dislocation b) a $-\frac{1}{2}$ disclination (note, the disclination core in the bottom configuration is in the vicinity of the elementary dislocation) c) a $+\frac{1}{2}$ disclination; solid dots correspond to elementary dislocations and $\pm$ signs correspond to disclination cores of strength $\frac{1}{2}$ \label{figdefects}}
\end{figure}

\subsection{Defects in the two-dimensional smectic-A mesophase} \label{secdefects}

Defects in mesophase order result in diverse texture and material properties.  Only the types of defects observed through the two-dimensional simulation of the isotropic/smectic-A phase transition will be addressed.  These defects include edge dislocations and wedge disclinations, shown in \Fref{figdefects}.  These defects can exist with different layering structures, where one is more energetically favourable than the other.  Dislocations have an associated Burger's vector, $\bi{b}$, which describes the topological charge associated with it.  In a right-handed coordinate system, both dislocation structures in \Fref{figdefects}a have $b=+1$.  Thus the sign of the Burger's vector corresponds to whether the dislocation begins ($+$) or ends ($-$) a set of layers, and the magnitude corresponds to the number of layers.

\begin{figure}[htp]
\begin{center}
\includegraphics[width=6in]{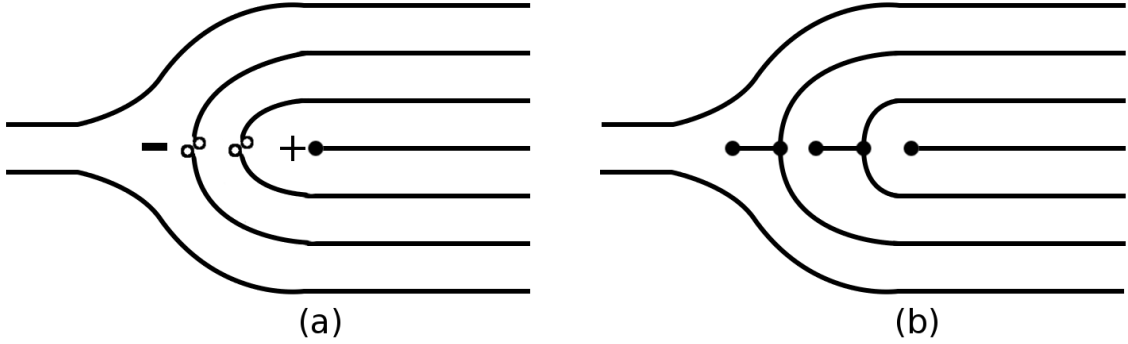} 
\end{center}

\caption{a) illustration of a lone $\pm\frac{1}{2}$ disclination pair interaction that forms a split-core dislocation where plus and minus signs correspond to $\pm\frac{1}{2}$ disclinations, solid circles correspond to elementary dislocations, and hollow circles correspond to dislocation content of an attached curvature wall (based on Figure 2a from \cite{Lejcek1990}) b) illustration of topologically equivalent array of elementary dislocations ; $b=+5$ for both of these defect structures. \label{figscdschem}}
\end{figure}

Many experimentally observed smectic defects such as twist dislocations and the diverse set of focal conic domains cannot be captured using two-dimensional simulation of a full order parameter model.  One type of composite defect is observed, the split-core dislocation or giant dislocation \cite{Kleman1982}, shown in \Fref{figscdschem}a.  This type of dislocation typically has a Burger's vector greater than one, where the dislocation core is split into a pair of $\pm\frac{1}{2}$ disclinations.  A single edge dislocation of Burger's vector greater than one is energetically unstable given the resistance of the layers to expansion/compression \cite{Pismen2006}.  Although, as shown in \Fref{figscdschem}b, an array of elementary dislocations can result in a topologically equivalent defect structure.

Referring to \Fref{figscdschem}a, a curvature wall is attached to the interacting disclinations and relieves dilation/compression forces resulting from the disclination cores \cite{Kleman1982,Lejcek1990}.  Curvature walls separate two disoriented smectic-A domains where the disorientation of the two domains is weak \cite{Kleman1982}.  They belong to a class of symmetric planar wall structures observed in the smectic-A mesophase which are outlined in \cite{Kleman1982}.

Little progress has been made in understanding the role of curvature walls in the dynamics of interacting smectic disclinations at the mesoscopic level.  Considering their key roles in the composition of the rich set of curvature defects resulting from the lamellar order of smectics \cite{Kleman1982}, there is motivation for studies focusing on the dynamics of these processes at the length and time scales at which they occur.  Theoretical approaches, in particular \cite{Lejcek1990}, have provided insight into the general structure of these wall defects.  Experimental observations have, at a macroscopic level, provided insight into the role of these wall defects in the coarsening process of similar lamellar systems \cite{Seul1992,Seul1992a} and more recently their role in FCD defects in smectics \cite{Blanc2000}.  In this context, an approach based upon modelling and simulation is appropriate for accessing these phenomena at the length and time scales of interest.

\subsection{Model and simulation approach} \label{secmodel}

A Landau-de Gennes type model for the first order isotropic/smectic-A phase transition is used that was initially presented by Mukherjee, Pleiner, and Brand \cite{deGennes1995,Mukherjee2001} and later extended by adding nematic elastic terms \cite{Brand2001,Mukherjee2002a,Biscari2007}:
\begin{eqnarray} \label{free_en}
f_b =&  \frac{1}{2} a \left( \bi{Q}:\bi{Q} \right) - \frac{1}{3} b \left( \bi{Q} \cdot \bi{Q} \right) : \bi{Q}+ \frac{1}{4} c \left( \bi{Q}:\bi{Q} \right)^2 + \alpha \left| \Psi \right|^2 + \beta \left| \Psi \right|^4  \nonumber\\
&- \frac{1}{2} \delta \left( \bi{Q}:\bi{Q} \right) \left| \Psi \right|^2 - \frac{1}{2} e \bi{Q} : \left(\bi{\nabla} \Psi \bi{\nabla} \Psi \right) \nonumber\\
&+ \frac{1}{2} l_1 \left( \bi{\nabla} \bi{Q} \vdots \bi{\nabla} \bi{Q} \right) +\frac{1}{2} b_1 \left| \bi{\nabla} \Psi  \right|^2 +\frac{1}{4} b_2 \left| \bi{\nabla}^2 \Psi  \right|^2  \nonumber\\
a  = & a_0 \left( T - T_{NI}\right) ; \alpha  =  \alpha_0 \left( T - T_{AI}\right)
\end{eqnarray}
where terms 1-5 are the bulk contributions to the free energy, terms 6-7 are couplings of nematic and smectic order; both the bulk order and coupling of the nematic director and smectic density-wave vector, respectively.  Terms 8-10 are the nematic and smectic elastic contributions to the free energy, respectively.  The order parameters are defined in \eref{eqnem_order_param}-\eref{eqsmec_order_param}, $T$ is temperature, $T_{NI}$/$T_{AI}$ are the hypothetical second order transition temperatures for nematic/smectic-A phase transitions (see \cite{Coles1979a}), and the remaining constants are phenomenological parameters.

\begin{figure}[htp]
\includegraphics[width=6in]{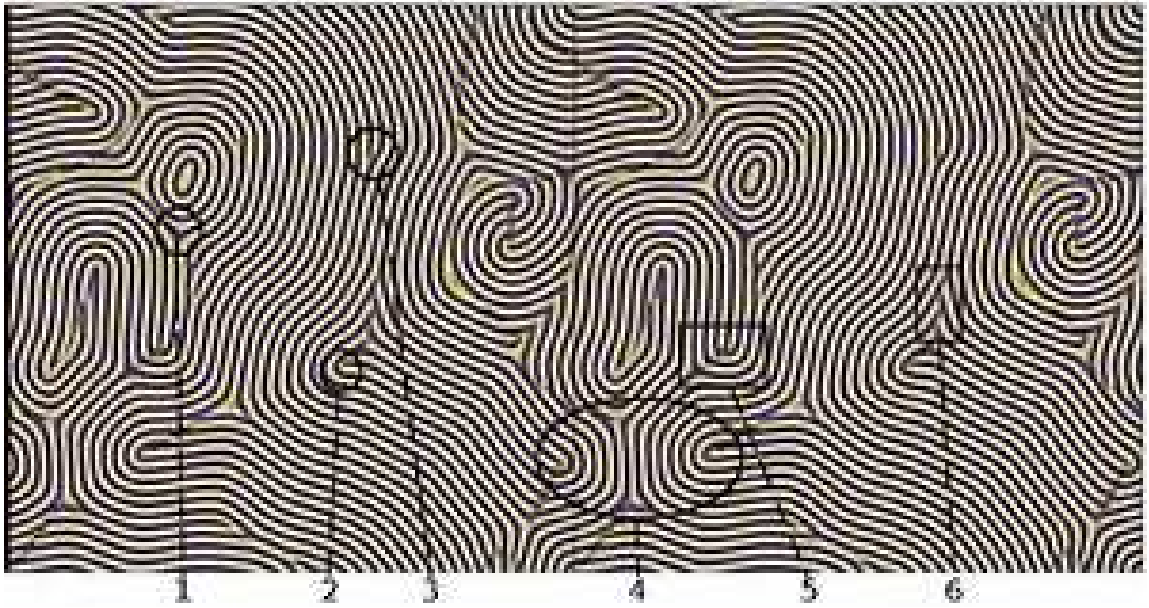} 
\caption{Sample computational domain where the surface corresponds to $Re(\Psi)$, the dimensionless density modulation, the contour lines are guides for the eye to show defects in smectic order, and periodic boundary conditions are shown by the fine vertical centreline; the selections outline different types of observed defects (see \Sref{secdefects} for more details): 1) a $-\frac{1}{2}$ disclination 2) a $+\frac{1}{2}$ disclination 3) an elementary dislocation 4) an angular quadrupole configuration of disclinations (see \Sref{secquad}) 5) a split-core dislocation 6) array of elementary dislocations; the phenomenological coefficients used in this simulation were $T=330K$, $T_{NI}=322.85K$, $T_{AI}=330.5K$, $a_0= 2\times10^5\frac{J}{m^3 K}$, $b=2.823\times10^7\frac{J}{m^3}$, $c=1.972\times10^7\frac{J}{m^3}$, $\alpha_0=1.903\times10^6\frac{J}{m^3 K}$, $\beta=3.956\times10^8\frac{J}{m^3}$, $\delta=9.792\times10^6\frac{J}{m^3}$, $e=1.938\times10^{-11}pN$, $l_1=1\times10^{-12}pN$, $b_1=1\times10^{-12}pN$, $b_2=3.334\times10^{-30}Jm$, and the ratio of the rotational and diffusional viscosities $\frac{\mu_S}{\mu_N}=25$.  The layer spacing is approximately $4nm$ which applies to all figures presenting simulation results in this work. \label{figbulk}}
\end{figure}

There are four aspects of this model which are key in the characterization of smectic-A order: i) use of full tensorial and complex order parameters, ii) inclusion of a second-order derivative of the complex order parameter, iii) inclusion of a nematic elastic term, and iv) imposition of the layer spacing through coupling with orientational order (term 7).  Clearly, the use of the full order parameters \eref{eqnem_order_param}-\eref{eqsmec_order_param} characterising orientational and translational order capture both continuum and phase transition phenomena.  Additionally, term 10 in the free energy density \eref{free_en} directly penalizes smectic layer curvature and is required both for stability and to capture the layer curvature contribution where the director deviates from the density-wave vector.  Preliminary simulations of two-dimensional droplet growth have shown that nematic elastic terms are also required due to nematic pre-ordering of the smectic front and to properly resolve disclination and elementary dislocation cores.  Past work has shown that inclusion of the $l_1$ and $l_2$ terms are necessary to resolve the isotropic/nematic interface, but we neglect the $l_2$ term in that our initial droplet configurations preclude the possibility to grow to length scales at which heterogeneity in the nematic/isotropic interface occurs \cite{Wincure2006}.  Finally, by using a formulation where the layer spacing is implicitly determined through the competition of coupling and elastic terms, simulation results show that this allows for a strong correlation to experimental observations (see \Fref{figbulk}). 

Material dependent coefficients are also an integral part of modelling and simulation using a phenomenological-type model.  The coefficients used in this work were determined by incorporating experimental data and the use of nonlinear programming, see \cite{Abukhdeir2007} for more details.  The use of physically realistic model parameters is clearly beneficial for comparison with experimental observations.

The Landau-Ginzburg time-dependent formulation is used to capture the kinetics of the phase transition.  Due to the higher order derivative term in the free energy functional a higher order functional derivative must be used.  The general form of the time-dependent formulation is as follows \cite{Barbero2000}:
\begin{equation}
\mu \frac{\partial \eta}{\partial t} = -\frac{\delta f_b}{\delta \eta} = \frac{\partial f_b}{\partial \eta} + \partial_i \left(\frac{\partial f_b}{\partial_i \eta}\right) - \partial_i \partial_j \left(\frac{\partial f_b}{\partial_i  \partial_j \eta}\right)
\end{equation}
where $\mu$ is viscosity, $\eta$ is the order parameter, and $f_b$ is the bulk free energy.  

A square computational domain  of $3.8\times 10^{-2} \mu m^2$ (approximately 50 layers) with full periodic boundary conditions was used to simulate bulk conditions.  The initial conditions for the simulation were a set of smectic droplets of equal diameter (approximately one smectic layer, $4nm$) randomly positioned and oriented throughout the domain.  The initial value for $S$, $\psi$, and the layer spacing were set to the homogeneous values for the corresponding temperature of $330K$.  The initial droplet number density was set at $3.287\times10^3 \mu m^{-2}$ and the random positioning algorithm was constrained to avoid initial impingement.  A sample of the simulation results is shown in \Fref{figbulk}.

\section{Results and discussion} \label{secresanddisc}

\Fref{figbulk} shows a sample texture at intermediate times in the coarsening process where the types of defects observed include elementary dislocations, $\pm \frac{1}{2}$ disclinations, curvature walls (of weak disorientation), and tilt walls (of strong disorientation) \cite{Kleman1982,Lejcek1990,Bowman1998,Pismen2006}.  Elementary dislocations, with a Burger's vector magnitude equalling $1$, are seen throughout the computational domain.  An important topological observation is that these smectic defects do not solely account for the total dislocation content of the domain.  Split-core dislocations, formed from pairs of smectic disclinations of opposite sign, account for the majority of the dislocation content.  This preference for split-core versus elementary dislocations is due to the energy associated with each, scaling proportional to $log(|b|)$ and $b^2$ \cite{Kleman1982}, respectively, where $b$ is the Burger's vector of the effective dislocation.

Further study of \Fref{figbulk} shows that curvature walls play an important role in the interactions of disclinations, and subsequently, split-core dislocations.  These split-core dislocations are composite defects and their dynamics are diverse.  The results of this work are presented sequentially, beginning with base disclination phenomena, followed by defect/texture kinetics in the coarsening process, and finally, with the dynamics of split-core dislocations.

\subsection{Disclination formation, movement, and basic pair annihilation} \label{secdiscform} 

The initial condition of randomly positioned and oriented droplets results in the formation of disclinations via the Kibble mechanism \cite{Kleman2003}.  The impingement of disoriented domains can be resolved in two general ways in a two-dimensional smectic-A system: disclination/curvature wall formation or tilt wall formation (arrays of elementary dislocations).  The formation of the disclination/curvature wall structure is energetically favourable, which is shown by their frequent occurrence as seen in \Fref{figbulk}.
 
\begin{figure}[htp]
\begin{center}\includegraphics[width=6in]{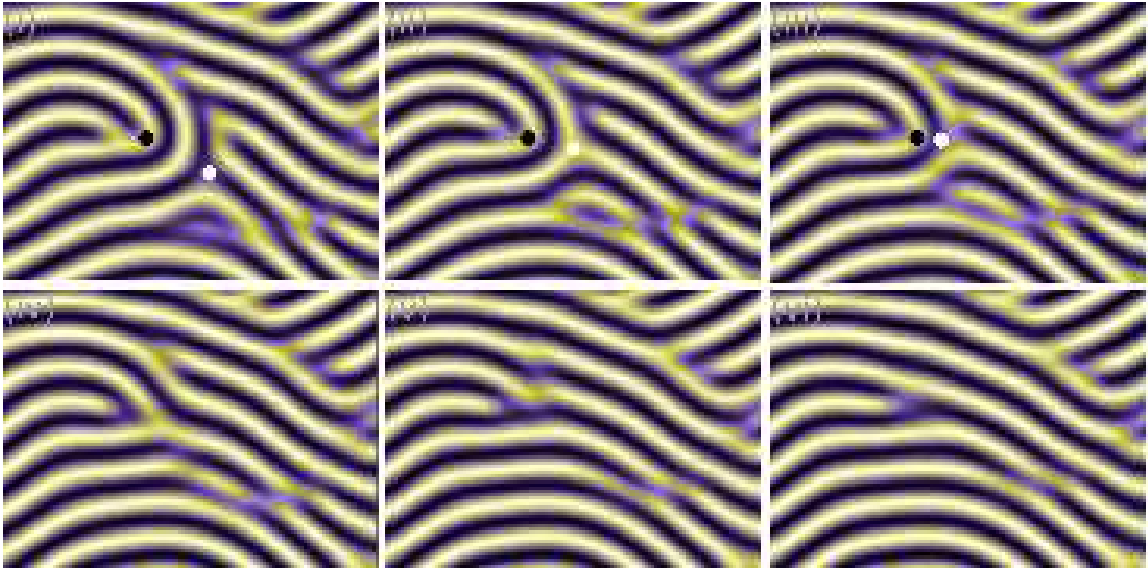}\end{center}
\caption{The annihilation of a $+\frac{1}{2}/-\frac{1}{2}$ disclination pair where: i) $t=0$ (relative), the initial configuration where the effective Burger's vector of the local area is $b=-1$ and the $+\frac{1}{2}$ disclination has just emitted a $-1$ dislocation ii) $t=500$, the $-\frac{1}{2}$ disclination absorbs a $+1$ dislocation (above it) resulting a decrease of the separation distance of the pair by one layer iii) $t=750$, another $+1$ dislocation is absorbed by the $-\frac{1}{2}$ disclination decreasing the separation distance of the pair by one layer iv) $t=1000$, the disclination pair annihilates emitting a $-1$ dislocation v) $t=1500$, the emitted dislocation annihilates with a $+1$ dislocation in its glide plane vi) $t=2000$, the post-annihilation region conserves the overall Burger's vector of $b=-1$;  black/white dots are guides for the eye to show $+\frac{1}{2}$/$-\frac{1}{2}$ disclinations.\label{disc_anni}}
\end{figure}

Following the growth period, defects begin to interact and are set in motion as a result of these forces.  The motion of elementary dislocations is well studied \cite{Kroupa1967,Pershan1975}, but the dynamics of smectic disclinations are understood only from a topological point of view.  Disclination motion is observed to require the absorption and emission of elementary dislocations.  The role of attached curvature walls, which relieve layer dilation/compression in the vicinity of the disclination core, will be discussed later in \Sref{seccurvwall1}.  Referring to \Fref{disc_anni}, the annihilation of two oppositely charged disclinations is limited by the topological constraints imposed by the smectic layers.  The movement of each disclination involves the absorption or emission of an elementary dislocation and a transition of the core of the disclination.  This process is seen for both types of disclinations in \Fref{disc_anni}, consistent with the predictions of topology theory and confirming the general coarsening mechanism proposed by Harrison et al \cite{Harrison2002}.

The annihilation of a disclination pair results in the emission of similarly charged elementary dislocations (resulting from disclination movement).  This relationship implies that the annihilation rates of each set of defects should differ, with that of elementary dislocations being the slower of the two.  Additionally, the total dislocation content of the domain must include the sum of elementary dislocations and dislocations associated with curvature walls (\Fref{figscdschem}a).  This dislocation content, which is directly related to the existence and interaction of disclinations, is accounted for by including defects in smectic order contained by curvature walls (see \Fref{figscdschem}a).  Following the disclination annihilation shown in the third frame of \Fref{disc_anni}, the resulting elementary dislocations emitted from the disclination annihilation process are free to interact. This process is shown in the last three frames of \Fref{disc_anni} where one of the emitted elementary dislocations annihilates with an oppositely charged neighbour.

\subsection{Defect/texture kinetics} \label{seckinetics}

Following the growth regime, multi-body interactions of defects dominate the texture dynamics.  The approximate self-similarity of smectic texture evolution in time, an inherent characteristic of chaotic natural systems, can be characterized by a governing power law of the form $t^n$.  Due to the periodic nature of the computational domain, an infinite smectic continuum is only approximated.  As the characteristic length of the texture approaches the periodic length scale, the system is no longer self-organized and deviates from scale-independence.  In this intermediate regime, the system organizes into a meta-stable state where the resulting texture evolution stagnates.  

\begin{figure}[htp] 
\begin{center}
\includegraphics[width=6in]{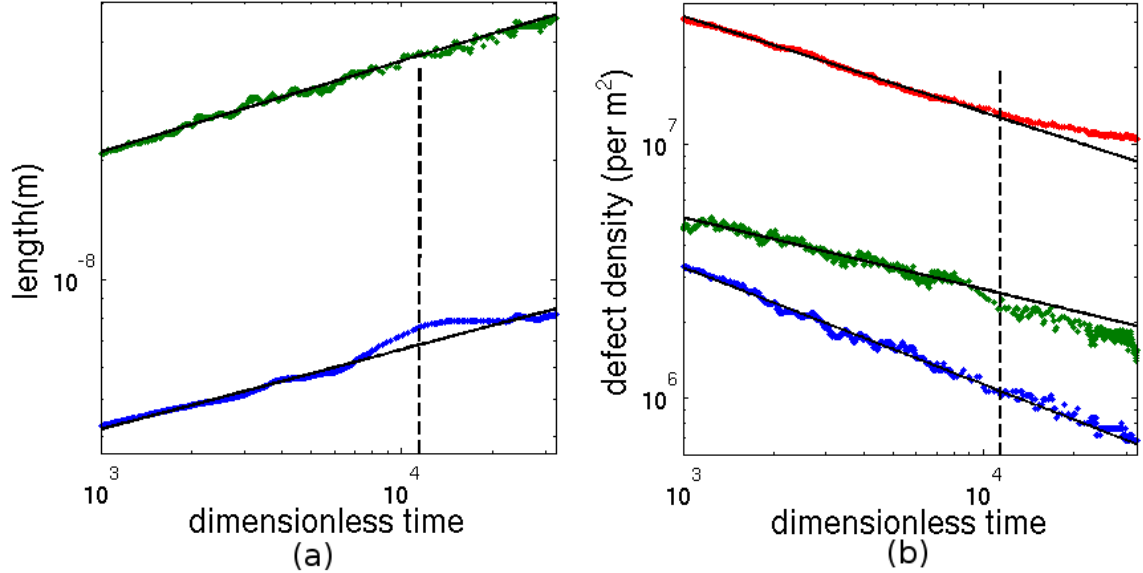}
\end{center}
\caption{a) (bottom) Orientational correlation length and (top) texture length scale ($\rho^{-0.5}$, where $\rho$ is the total density of disclinations) versus time; the fit curves are $t^{0.20}$ and $t^{0.23}$, respectively b) defect densities versus time for disclinations (bottom), elementary dislocations (middle), and total dislocation content (top); the fit curves are $t^{-0.46}$, $t^{-0.28}$. and $t^{-0.38}$, respectively. The dotted lines in both subfigures denote the transition away from the self-similar coarsening regime to what is referred to as the intermediate coarsening regime. \label{figcoarsening}}
\end{figure}

In order to accurately characterize the coarsening process, the orientational correlation length was determined directly from the tensor order parameter \eref{free_en} \cite{Denniston2001}.  The growth, coarsening, and meta-stable regimes were determined from this result, shown in \Fref{figcoarsening}a.  In addition, the texture length scale (the average separation distance between disclinations) was calculated using the disclination density, $\rho^{-\frac{1}{2}}$, and both quantities were found to exhibit power law behaviour with $n=\frac{1}{4}$.  This implies that the coarsening process is dominated by disclination interactions.  Furthermore, this shows that the disclination interactions deviate from those observed in the nematic phase which yield an exponent of $n=\frac{1}{2}$ for the texture length scale evolution \cite{Denniston2001}.  These results are in good agreement with previous experimental work \cite{Harrison2000,Harrison2002} and the deviation from the nematic phase is attributed the topological constraints of disclination movements and interactions, which was shown in \Sref{secdiscform}.

The density of disclinations, elementary dislocations, and total dislocation content was determined and is shown in \Fref{figcoarsening}b.  The disclination density is found to be governed by $t^{-\frac{1}{2}}$ power law, agreeing with experimental observations \cite{Harrison2000,Harrison2002}.  Additionally, both dislocation densities are found to obey a $t^{-\frac{1}{3}}$ power law, differing from that of the disclinations.  This difference is qualitatively intuitive from observations in the previous section, that annihilation of disclinations can occur without changing the net dislocation content.  This result is also supported by past experimental observations which found that dislocations annihilate at a slower rate than disclinations \cite{Harrison2002}.

Past modelling and simulation studies of other lamellar pattern-forming systems have shown that dislocation annihilation rates obey a power law where $n=-\frac{1}{3}$ \cite{Boyer2002}.  The effects of the incorporation of fluctuations were also studied showing that they play a role at intermediate time scales in the coarsening process and at quench depths approaching the bulk transition temperature \cite{Boyer2002}.  Thus the incorporation fluctuations would not have a substantial effect on the coarsening results found here in that the quench depth is below the lower stability limit of the isotropic phase and the orientational correlation length was used to determine the self-similar coarsening regime (\Fref{figcoarsening}a).

In the present model, couplings between the phase ordering dynamics and momentum balance equation are also neglected. This excludes convection effects due to defect-defect interactions and re-orientation induced flow.  For example, it is known that interaction between two disclinations creates asymmetric vortex flows that renormalize the rotational viscosity \cite{Blanc2005}.  Nevertheless, in the present work which is focused on texturing statistics of smectic-A materials, the degree of agreement between the presented simulation results and experimental observations strongly suggests that neglecting convection does not result in deviation from actual texturing statistics.

\subsection{Role of curvature walls in disclination interaction} \label{seccurvwall1}

At intermediate times (see \Fref{figcoarsening}a) the coarsening process deviates from the previously determined power law behaviour as the texture length scale approaches the imposed periodic length.  At this point, the number of interacting defects becomes small and basic interactions of smectic disclinations are more easily observed.  These interactions provide insight into the structure of many physical systems that related to smectic-A and other lamellar ordered systems.

\begin{figure} 
\begin{center}\includegraphics[width=6in]{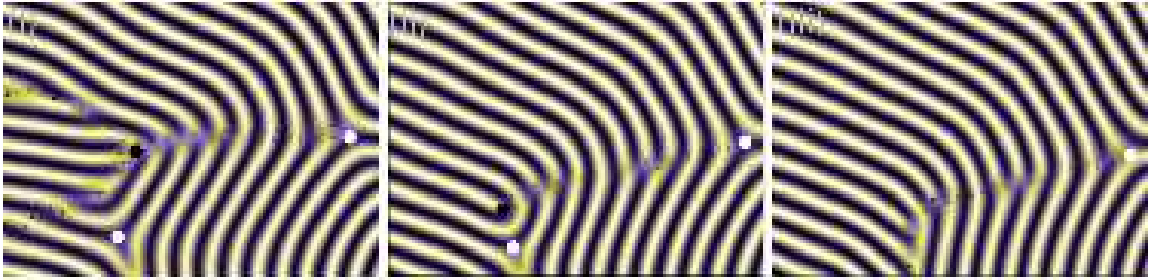}\end{center}
\caption{The annihilation of a pair of oppositely charged disclinations facilitated by the presence of a curvature wall: (i) $t=0$ (relative), the initial configuration where the $+\frac{1}{2}$ disclination begins the emission process an elementary dislocation into an attached curvature wall (ii) $t=5000$, the $+\frac{1}{2}$ disclination approaches a nearby $-\frac{1}{2}$ disclination through the absorption and emission of elementary dislocations into an attached curvature wall, forming a split-core dislocation structure (iii) $t=10000$, the disclination pair annihilates emitting their total dislocation content into an attached curvature wall; black/white dots are guides for the eye to show $+\frac{1}{2}$/$-\frac{1}{2}$ disclinations and contour lines outline defects in the smectic order. \label{figdiscinteract}}
\end{figure}

The texture structures that are found to form at intermediate times clearly show the importance of curvature walls to the movement disclinations and elementary dislocations.  Disclinations of opposite signs tend to pair and form split-core dislocations of high Burger's vector, shown in \Fref{figdiscinteract}.  Interactions of these split-core dislocations, which are essentially quadrupolar disclination interactions, have been observed experimentally in other mesomorphic systems including cholesterics \cite{Kleman1982}, nematics under an applied field \cite{Hudson1991}, liquid crystalline polymer melts \cite{Wood1986}, and most recently cylindrical block-copolymer films \cite{Harrison2000,Harrison2002}.  These recent experimental observations illuminate the importance of understanding these complex interactions.

Attached curvature walls play a key role in the dynamics of split-core dislocations.  \Fref{figdiscinteract} shows the formation of an unstable split-core dislocation which eventually disassociates into an array of elementary dislocations.  The role of attached curvature walls as a low energy pathway for the movement, emission, and absorption of elementary dislocations is clearly shown.  Referring to \Fref{figdiscinteract}, the curvature wall on the right side of the $+\frac{1}{2}$ disclination begins with no net Burger's vector and transitions to a state with $b>0$.  The curvature wall to the left acts as a low-energy pathway for elementary dislocation movement to the $+\frac{1}{2}$ disclination.  Through the absorption and emission process, the $+\frac{1}{2}$ disclination approaches a neighbouring oppositely charged disclination and the pair annihilates.  In this case the split-core dislocation structure was transient and unstable.

\subsection{Quadrupolar interactions of disclinations} \label{secquad}

\begin{figure}[htp] 
\begin{center}\includegraphics[width=6in]{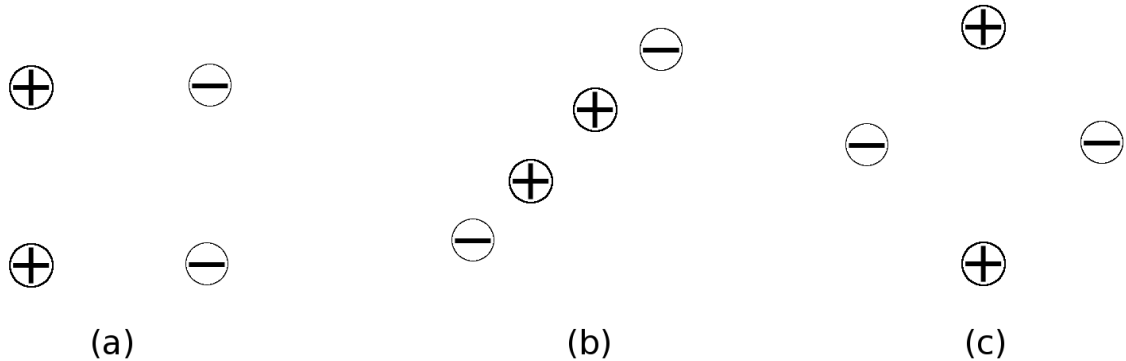}\end{center}
\caption{Schematic of the three different configurations of quadrupolar interactions: a) effective dipole, b) linear quadrupole, and c) angular quadrupole; adapted from \cite{Hudson1991}. \label{fighudson}}
\end{figure}

Interactions between disclinations are complex due to the topological constraints imposed on their movements by lamellar order.  Experimental observations have shown that these interactions are essentially quadrupolar interactions of disclinations \cite{Harrison2002}.  Utilising the approach by Hudson and Thomas for the characterization of quadrupolar disclination interactions in an applied field \cite{Hudson1991}, different configurations of quadrupolar interactions are shown in \Fref{fighudson}.  The angular quadrupole is the lowest energy formation in that it results in the greatest amount of shielding of like-signed defects \cite{Hudson1991}.  This topological configuration is also favourable in the smectic-A mesophase because it allows for a highly localized disoriented domain which is resolved almost entirely by curvature.  This explains the preference for this configuration observed in \Fref{figbulk}.  The majority of disclination interactions at the intermediate time-scale occur in an angular quadrupole configuration.  The idealized version of this structure would be composed of split-core dislocations of equal Burger's vector magnitude and would annihilate isomorphically.

\begin{figure} 
\begin{center}\includegraphics[width=6in]{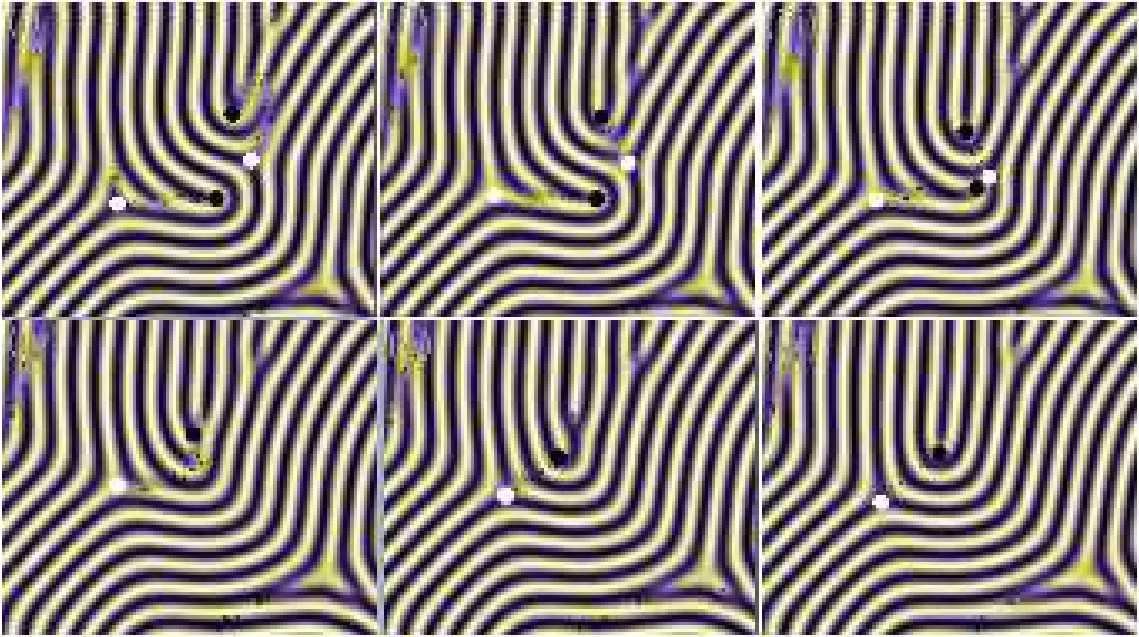}\end{center}
\caption{A pair of split-core dislocations of similar sign but differing Burger's vector magnitude ($b=2$ for lower and $b=4$ for upper) interact in a mixed angular quadrupole/effective dipole configuration (see \Fref{fighudson}) i) $t=0$ (relative), the initial configuration where both split-core dislocations are of the same sign (the interaction must result in addition/consolidation) ii) $t=5000$, the $+\frac{1}{2}$ disclination in the upper split-core dislocation emits an elementary dislocation allowing glide of the pair towards the $-\frac{1}{2}$ disclination of the lower split-core dislocation iii) $t=10000$, annihilation of the inner pair of oppositely charged disclinations occurs joining the respective curvature walls iv) $t=11667$, the resulting dislocation structure emitted by the annihilated disclinations is absorbed into the curvature wall of the remaining split-core dislocation, conserving the total Burger's vector ($b=6$) v) $t=13333$, an initially emitted elementary dislocation climbs toward the $+\frac{1}{2}$ disclination  vi) $t=25000$, the additive interaction of the two split-core dislocations is complete resulting a single split-core dislocation with Burger's vector equalling the sum of the original two; black/white dots are guides for the eye to show $+\frac{1}{2}$/$-\frac{1}{2}$ disclinations and contour lines outline defects in the smectic order.  \label{figangquadanni}}
\end{figure}

Interactions between split-core dislocations of different magnitudes and in non-ideal configurations are more likely.  Referring to \Fref{figbulk}, all of the angular quadrupole configurations are composed of non-equal pairs of split-core dislocations.  When such a mismatch occurs, a fully isomorphic annihilation process is not possible.  \Fref{figangquadanni} shows a disclination quadrupole in a mixed angular quadrupole/effective dipole configuration.  This interaction shows a mechanism absent from interactions of elementary dislocations, where two split-core dislocations partially annihilate.  This can be thought of as an addition of two split-core dislocations where the inner disclination pair annihilates and the by-product elementary dislocations resulting from their movement are absorbed by the split-core dislocation that remains.  This process is found to strongly involve the curvature walls associated with respective split-core dislocations.  The resulting texture has an equivalent Burger's vector of the sum of the two original split-core dislocations.
 
The addition mechanism for split-core dislocations is determined from these simulation results, but annihilation was not observed.  Candidates for such an interaction are found to form low-energy angular quadrupole configurations (\Fref{figbulk}) and are meta-stable at the time-scales accessible.  Using insight from the results of this work and previous experimental observations \cite{Harrison2002,Hudson1991}, the mechanism for this annihilation process of a pair of split-core dislocations can be deduced.  An iterative process where single elementary dislocations are emitted and then absorbed by a pair of split-core dislocations in an angular quadrupole configuration would result in an energy barrier to the process being on the order of the core energy of an elementary dislocation.  This proposed mechanism is compatible with the constraints observed both here through simulation and experimentally \cite{Harrison2002}.  This would satisfy the basis for the coarsening power law exponent, that disclination interactions are limited by the movement of elementary dislocations on the order of the separation distance (of the disclinations).

\subsection{Curvature wall instability} \label{seccurvwall2}

\begin{figure} 
\begin{center}\includegraphics[width=6in]{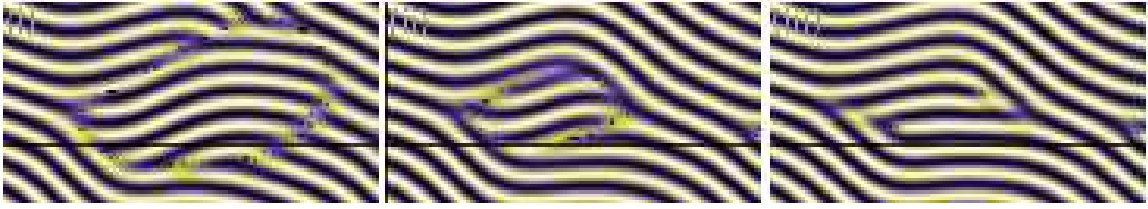}\end{center}
\caption{A decoupled curvature wall self-annihilates: i) $t=0$ (relative), in its initial state after detaching from a disclination ii) $t=10000$, shrinking isomorphically iii) $t=20000$, after fully annihilating an array of elementary dislocations remains; contour lines outline defects in the smectic order and periodic boundary conditions are shown by the fine horizontal line.  \label{figcurvwall}}
\end{figure}

The role of curvature walls in the context of disclination interaction has been discussed, but the dynamics of unattached curvature walls has not.  The behaviour of this wall defect removed of a disclination population to meta-stabilize it is shown in \Fref{figcurvwall}.  As a pair of oppositely charged disclinations form a split-core dislocation (from \Fref{figdiscinteract}), one of the curvature walls becomes unattached and isolates a disoriented domain.  This domain structure is similar to that seen in the angular quadrupole configuration of disclinations where the disoriented domain is now resolved topologically without disclinations.  This domain is unstable in its initial configuration and its total area is minimized allowing a local relaxation of the layer curvature.  The annihilation process is iterative where pairs, along the axis parallel to the smectic layers, of dislocations associated with the curvature wall approach and annihilate expelling any elementary dislocation content that was previously absorbed.  The disoriented domain decays in an isomorphic process, conserving the topological charge.  A meta-stable array of elementary dislocations isolate the remaining area of the disoriented domain which is stabilized by a texture-imposed localized dilation of the layers. 

This process was not observed in previous experimental work \cite{Harrison2002} due to the length scales at which such interactions occur and the difficulty of identifying transient curvature walls.  These results suggests that in addition interactions of split-core dislocations, isolated curvature wall loops have a role in the texturing process.

\section{Conclusions}

The growth and coarsening process of the two-dimensional first-order isotropic/smectic-A mesophase transition was studied, texture growth laws determined, and underlying defect interactions explained.  The role of smectic disclinations was shown to be key in the coarsening process and their dynamics strongly deviating from those in the nematic phase.  These interactions are summarized:\begin{itemize}
\item Topologically constrained smectic disclination interaction is limited by the absorption and emission of elementary dislocations (\Fref{disc_anni}) and this process does not necessarily result in a change in total dislocation content (\Fref{figcoarsening}b).
\item The coarsening process is governed by the annihilation of disclinations where the orientational correlation length obeys a power law with exponent $n=\frac{1}{4}$ (\Fref{figcoarsening}a); defect annihilation power law behaviour for disclinations and dislocations differs, where $n=-\frac{1}{3}$ for both (\Fref{figcoarsening}b).
\item Curvature walls have a fundamental role in disclination interactions and following the multi-body interaction regime, split-core dislocations form from disclination pairs of opposite sign and their corresponding curvature walls (\Fref{figdiscinteract}).
\item Base split-core dislocation interactions behave as quadrupolar configurations of disclinations which accurately predict their dynamics and meta-stable angular quadrupole configurations (Figures \ref{figbulk} and \ref{figangquadanni}).\item Curvature walls are meta-stable defect structures and if detached from stabilizing disclinations self-annihilate in an isomorphic process (\Fref{figcurvwall}).
\end{itemize}The phenomena observed through this modelling and simulation study, such as angular quadrupoles of disclinations, may be responsible for many of the unique material properties of materials exhibiting lamellar ordering \cite{Wood1986}.

Past theoretical and experimental studies have provided insight into much of the phenomena presented, but numerical simulation of finer-grained models is currently the only way to study these phenomena further.  The simulation results presented show that the use of a high-order phenomenological model and experimentally based model parameters results in a substantially more complete reproduction of the physical smectic-A system.  Though current computational resources restrict this initial work to two-dimensions, these results show a strong correlation to experimental observations of some of the complex curvature defects found in smectic liquid crystals.  These promising results show that three-dimensional simulation of this model could be used to study the formation and dynamics of focal-conic defects on length and time scales currently unaccessible via experimental study.

From the viewpoint of industrial applications, the two-dimensional smectic-A system shows a great deal of promise for use in cylindrical block copolymer systems.  Cutting-edge experimental work on enhancement of the photolithographic process using block copolymer films \cite{Ruiz2007,Stoykovich2007,Black2007} could benefit from the presented modelling and simulation approach in addition to the generation of material specific model parameters \cite{Abukhdeir2007}, incorporation of edge-effects \cite{Abukhdeir2008}, and bulk field terms.

\ack
This work was partially funded by the Natural Science and Engineering Research Council of Canada.

\section*{References}

\bibliography{references}

\end{document}